\documentclass[journal,twoside,web]{ieeecolor2}
\usepackage{generic}
\usepackage{cite}
\usepackage{amsmath,amssymb,amsfonts}
\usepackage{algorithmic}
\usepackage{graphicx}
\usepackage{textcomp}
\usepackage{booktabs}

\def\BibTeX{{\rm B\kern-.05em{\sc i\kern-.025em b}\kern-.08em
    T\kern-.1667em\lower.7ex\hbox{E}\kern-.125emX}}
\begin{document}
\title{Coronary Mask Guided Registration for Continuous Time 4D Cardiac CT Dataset Construction}
\author{Yuang Wang, Shuo Wang, Changyu Chen, Dufan Wu, Pengfei Jin, Yunqiang An, Yang Gao, Bin Lu, Dongrui Dai, Muge Du, Yan Yan, Dong Li, Liang Li, Li Zhang, Zhiqiang Chen
\thanks{Yuang Wang, Shuo Wang, Changyu Chen, Dongrui Dai, Muge Du, Liang Li, Li Zhang and Zhiqiang Chen are (were) with the Department of Engineering Physics, Tsinghua University, Beijing 100084, China.}
\thanks{Dufan Wu is with the Department of Radiology, The Ohio State University, 2050 Kenny Rd, Columbus, OH
43221, Ohio, USA.}
\thanks{Pengfei Jin is with Center for Advanced Medical Computing and Analysis, Massachusetts General Hospital
and Harvard Medical School, 399 Revolution Dr, Somerville, 02145, Massachusetts, USA.}
\thanks{YunQiang An, Yang Gao, Bin Lu are with the Department of Radiology, Fuwai Hospital, Chinese Academy of Medical Sciences \& Peking Union Medical College, Beijing, China.}
\thanks{Yan Yan and Dong Li are with the Department of Radiology, Tianjin Medical University General Hospital, Tianjin, China.}
\thanks{Send correspondence to Li Zhang (zli@mail.tsinghua.edu.cn) and Zhiqiang Chen (czq@mail.tsinghua.edu.cn) }
}

\maketitle

\begin{abstract}
\textit{Objective:} Clinical cardiac CT multiphase reconstructions generally provide acceptable image quality in end-diastole (ED) or end-systole (ES) phases, but in other phases may exhibit motion artifacts, especially in the right coronary artery (RCA). This limits ground-truth availability in 4D cardiac CT imaging research. We aim to construct a 4D cardiac CT dataset that is generally suitable to serve as pseudo ground truth. 
\textit{Methods:} We propose Coronary Mask Guided Registration (CMGR) to produce a motion-preserved, artifact-reduced, and continuous-time 4D cardiac CT sequence from the clinical multiphase reconstruction of each patient. For artifact reduction, CMGR uses the ED or ES phase as the reference phase and warps the reference volume with deformation fields to produce the sequence. For motion preservation, CMGR registers the reference phase to each non-reference phase of the multiphase reconstruction. To capture the motion of both the RCA and other cardiac structures in each registration, CMGR regularizes RCA masks and incorporates them into image-domain registration. Time-continuity is achieved by interpolating the deformation fields for non-reference phases to arbitrary times.
\textit{Results:} CMGR outperformed representative image-domain registration methods in capturing RCA motion and providing reasonable RCA shape, and showed competitive performance in capturing whole-heart motion. Additionally, CMGR reduced motion artifacts from clinical multiphase reconstructions, and intermediate CMGR frames generally provided plausible transitions between discrete cardiac phases.
\textit{Conclusion:} CMGR provides an effective approach for constructing continuous-time 4D cardiac CT datasets. 
\textit{Significance:} The dataset can be used in system design simulations and in reconstruction algorithm development, thereby facilitating advances in cardiac CT imaging.
\end{abstract}

\begin{IEEEkeywords}
4D cardiac CT, registration, right coronary artery
\end{IEEEkeywords}

\section{Introduction}
\label{sec:introduction}
\IEEEPARstart{C}{ardiac} computed tomography (CT) is an important noninvasive imaging technique for visualizing cardiac anatomy and assessing cardiovascular diseases. However, real cardiac motion remains inaccessible. Clinical multiphase reconstructions generally provide acceptable image quality in phases close to end-diastole (ED) or end-systole (ES), but in other phases may exhibit severe motion artifacts, particularly in the right coronary artery (RCA) region. 

This poses a common challenge in constructing 4D cardiac CT datasets that serve as pseudo ground truth for 4D cardiac CT imaging research. Pseudo ground truth is typically required in simulation studies for CT system design, quantitative evaluation of image reconstruction methods, and supervised training of deep learning-based reconstruction approaches. 

Previous studies typically constructed pseudo ground truth using four categories of strategies. The first category~\cite{yan2025end} directly uses clinical multiphase reconstructions as pseudo ground truth; however, many phases exhibit motion artifacts. The second category~\cite{lossau2019motion,maier2021deep} manually designs deformation fields around the RCA centerline and applies them to well-reconstructed volumes; however, such deformation fields are generally restricted to RCA and difficult to propagate to other cardiac structures. The third category employs digital phantoms such as the 4D extended cardiac-torso (XCAT) phantom~\cite{segars20104d}, which provide continuous-time cardiac motion but may simplify anatomical details compared with real patients. The fourth category~\cite{deng2023tt} extracts cardiac motion from XCAT and transfers it to well-reconstructed clinical volumes; however, cardiac motion diversity may be limited by XCAT.

Since clinical multiphase reconstructions often suffer from motion artifacts but provide patient-specific cardiac anatomy and cardiac motion, one may consider reducing these artifacts using established motion-compensated reconstruction methods and subsequently using the resulting sequences as pseudo ground truth. However, reconstruction methods are generally unsuitable for this purpose. Most classical motion estimation and motion-compensated reconstruction (ME-MCR) methods~\cite{kim2015cardiac,bhagalia2012nonrigid} and motion-aware networks~\cite{yan2025end,deng2026marvel} require raw projection data, which are often unavailable due to proprietary restrictions. This limits their applicability for constructing large-scale datasets. Reprojection approaches~\cite{hahn2017motion,maier2021deep} and post-processing networks~\cite{lossau2019motion,jung2020deep} can operate directly on clinical multiphase reconstructions and reduce motion artifacts. However, they typically produce the same number of cardiac phases as the multiphase reconstruction. The temporal resolution of such outputs may be insufficient for developing cardiac CT systems~\cite{chen2025static} and reconstruction algorithms that aim to achieve substantially higher temporal resolution than clinical multiphase reconstructions.

In this work, we propose Coronary Mask Guided Registration (CMGR) to produce a motion-preserved, artifact-reduced, and continuous-time 4D cardiac CT sequence directly from the clinical multiphase reconstruction of each patient. For artifact reduction, CMGR uses the cardiac phase with minimal motion artifacts as the reference phase and warps the reference volume with deformation fields to produce the sequence. For motion preservation, CMGR registers the reference phase to each non-reference phase of the multiphase reconstruction to produce deformation fields. Time-continuity is achieved by interpolating deformation fields for non-reference phases to arbitrary times.

In registering the reference phase to each non-reference phase, directly applying existing registration methods may be suboptimal. Image-domain registration methods, including both conventional registration methods~\cite{rueckert1999nonrigid,ASHBURNER200795} and learning-based approaches~\cite{dalca2019unsupervised,tian2024unigradicon}, may fail to capture RCA motion in the non-reference phases where the RCA suffers from severe motion artifacts or exhibits large displacement relative to the reference phase. Centerline-focused registration methods~\cite{hadjiiski2014coronary} can provide acceptable motion estimation of the RCA centerline, but they do not explicitly preserve RCA shape and generally cannot capture chamber and aorta motions.

In each registration, to capture the motion of both the RCA and other cardiac structures, CMGR regularizes the RCA masks for both reference and non-reference phases and incorporates them into image-domain registration. Specifically, RCA masks are regularized for connected main trunks, reasonable shape, plausible centerlines and consistent branches. These masks are then incorporated into a proposed multi-resolution registration strategy for producing a deformation field that captures RCA motion and preserves RCA geometry. An additional deformation field that captures the remaining cardiac motion without distorting the RCA is subsequently produced using our proposed RCA protection strategy. Composing the two deformation fields yields a whole-heart deformation field that generally preserves cardiac motion from clinical multiphase reconstructions.

CMGR is compared with representative image-domain registration approaches. Through extensive experiments, CMGR demonstrates superior capability and stability in capturing RCA motion and providing reasonable RCA shape, and shows competitive performance in capturing whole-heart motion. We also show CMGR's capability in reducing motion artifacts from clinical multiphase reconstructions. In addition, we demonstrate that intermediate CMGR frames generally provide plausible transitions between discrete cardiac phases.

The main contributions are summarized as follows:
\begin{itemize}
\item{We propose CMGR for 4D cardiac CT dataset construction. To the best of our knowledge, CMGR is the first approach to produce motion-preserved, artifact-reduced, and continuous-time 4D cardiac CT sequences from clinical multiphase reconstructions without requiring raw projection data.}
\item{To the best of our knowledge, CMGR is the first registration approach that explicitly regularizes RCA masks and incorporates them into image-domain registration. This enables CMGR to capture the motion of both the RCA and other cardiac structures.}
\item{Through extensive experiments, we demonstrate CMGR's strong capability in capturing RCA motion and providing reasonable RCA shape, as well as its competitive performance in capturing whole-heart motion. We also show that CMGR is able to reduce motion artifacts from clinical multiphase reconstructions and that intermediate CMGR frames generally provide plausible transitions between discrete cardiac phases.}
\end{itemize}

A preliminary version of this work has been reported in~\cite{preliminary}.

\section{Methods}

\begin{figure*}[t]
    \centering
    \includegraphics[width=1\linewidth]{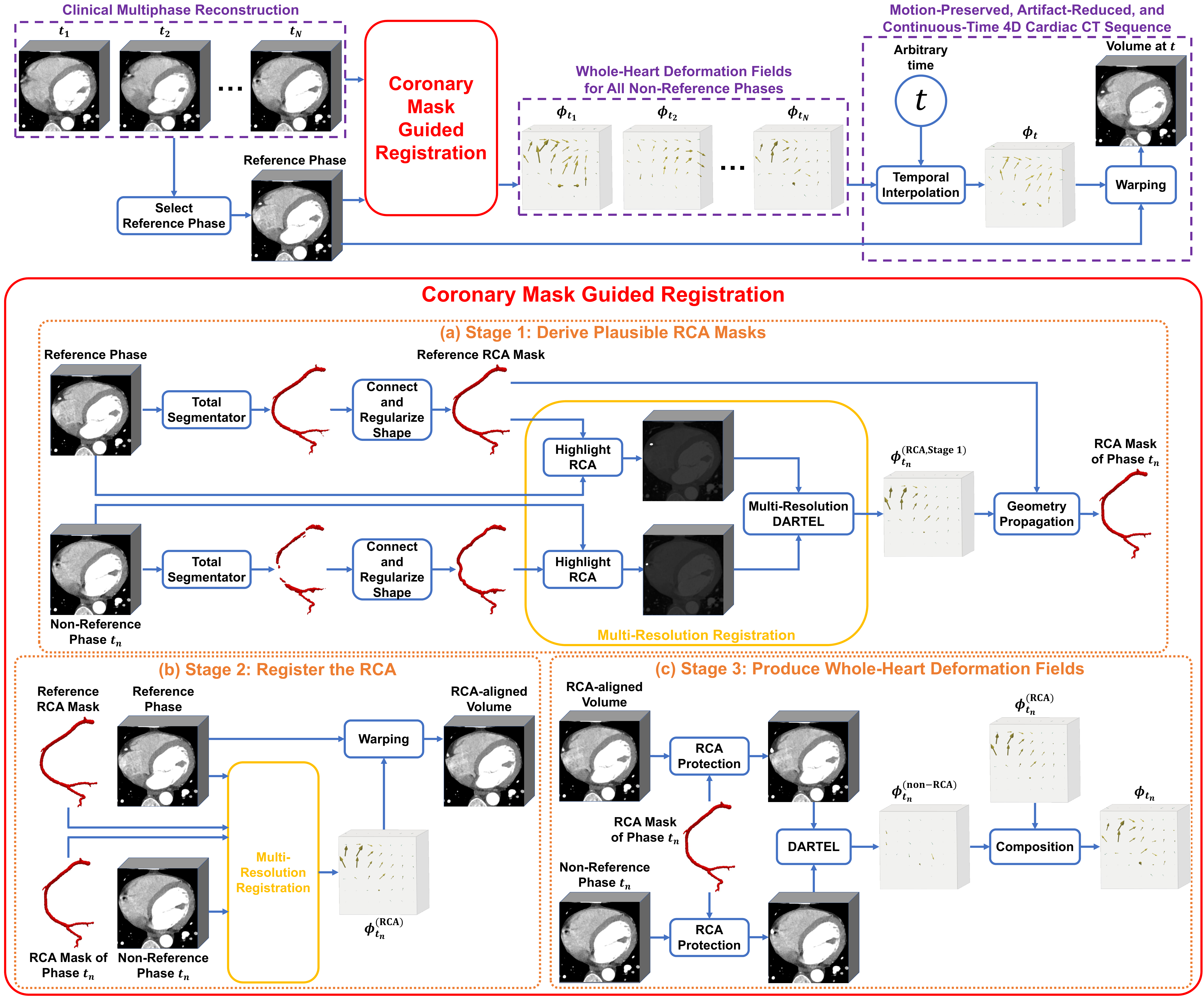}
    \caption{CMGR framework. CMGR is designed to produce a motion-preserved, artifact-reduced, and continuous-time 4D cardiac CT sequence from a clinical multiphase reconstruction. CMGR uses the cardiac phase with minimal motion artifacts as the reference phase and registers the reference phase to each non-reference phase. The two yellow boxes labeled “Multi-Resolution Registration” in Stages 1 and 2 represent the same registration strategy applied to different inputs. Specifically, the non-reference RCA mask in Stage 2 is the RCA mask produced after geometry propagation.} 
    \label{fig:method} 
    
\end{figure*}
In this section, we present our proposed method, CMGR. CMGR is designed to produce a motion-preserved, artifact-reduced, and continuous-time 4D cardiac CT sequence from the clinical multiphase reconstruction of each patient, thereby facilitating large-scale 4D cardiac CT dataset construction. The overall framework of CMGR is illustrated in Fig. \ref{fig:method}.

From the clinical multiphase reconstruction, CMGR uses the cardiac phase with minimal motion artifacts as the reference phase. The reference phase is typically near ED or ES. When ED and ES have comparable image quality, ED is preferred because the RCA position at ED is generally more central in the RCA motion trajectory.

CMGR registers the reference phase to each non-reference phase $t_n$ in the clinical multiphase reconstruction. This registration produces the deformation field  $\phi_{t_n}$ that captures the whole-heart motion. The deformation field $\phi_{t_n}$ is defined as the transformation that maps the spatial position $x$ at phase $t_n$ to the position $\phi_{t_n}(x)$ at the reference phase.
The whole-heart deformation fields  for all non-reference phases, together with the reference volume, are used to produce the continuous-time cardiac CT sequence. Specifically, the deformation field $\phi_t$ at an arbitrary time $t$ is obtained by interpolation:
\begin{equation}
\phi_{t}=\sum_{n=1}^{N}
B_n^{(1)}(t)
\phi_{t_n},
\label{eq:inter}
\end{equation}
where $N$ is the total number of phases in the multiphase reconstruction, and $B_n^{(1)}(t)$ is the $n$th first-order B-spline basis function at $t$. The continuous-time sequence $\hat{I}_t$ is obtained by
\begin{equation}
\hat{I}_t = I_{\text{ref}}\circ\phi_t,
\label{eq:warp}
\end{equation}
where $I_{\text{ref}}$ is the reference volume, and $\circ$ is the warping operation~\cite{jaderberg2015spatial}.
As $I_{\text{ref}}$ is always the warping source, the produced sequence $\hat{I}_t$ maintains the motion artifact level at all time points comparable to that of the reference phase.

When registering the reference phase to each non-reference phase, the main challenge lies in capturing the RCA motion and preserving the RCA geometry. This is because the RCA may exhibit large displacement or suffer from severe motion artifacts. To address this challenge, CMGR consists of three stages. In Stage 1, plausible RCA masks for both reference and non-reference phases are derived. Regularization procedures are introduced to alleviate the influence of motion artifacts on the RCA masks. In Stage 2, a deformation field that captures RCA motion and preserves RCA geometry is produced under the guidance of the RCA masks using a proposed multi-resolution registration strategy. In Stage 3, the remaining cardiac motion is captured without distorting the RCA, and the whole-heart deformation field $\phi_{t_n}$ is obtained. The details of these three stages are described in the following subsections.
\subsection{Stage 1: Derive Plausible RCA Masks}
Stage 1 of CMGR derives plausible RCA masks for reference and non-reference phases. The workflow is illustrated in Fig.~\ref{fig:method}(a). We initialize the RCA masks using TotalSegmentator~\cite{wasserthal2023totalsegmentator}. For the reference phase, the initial RCA mask generally exhibits reasonable geometry. However, for patients with anatomically thin or partially occluded RCA, the initial RCA mask may exhibit a discontinuous RCA main trunk. For non-reference phases, motion artifacts may substantially degrade segmentation quality. The initial RCA masks may exhibit discontinuous RCA main trunks, distorted shape, over-deformed centerlines, and missing or spurious branches.

To improve the geometric integrity of the initial RCA masks, we design regularization procedures including connecting the RCA main trunk of the mask, regularizing the RCA mask shape, and propagating RCA mask geometry of the reference phase to non-reference phases. The details of these procedures are described in the following subsubsections.

\subsubsection{Connect RCA Main Trunk}
The purpose here is to produce RCA masks with connected RCA main trunks. The proximal and distal endpoints of the RCA main trunk in the reference phase are used as patient-specific anchor points to locate RCA mask fragments that may need to be connected. For each initial RCA mask, we identify the fragments closest to the proximal and distal anchor points as the proximal fragment and distal fragment, respectively. If they are not connected, we connect them with the following procedures.

We determine the pair of boundary points that minimize the Euclidean distance between the proximal and distal fragments, and connect the two boundary points using the minimum-cost path on the cost map $C(x)$:
\begin{equation}
C\left(x\right)=\begin{cases}
\frac{\left(I\left(x\right)-\mu\right)^2}{2\sigma^2}-\ln p_{\text{Total}}\left(x\right), & \text{if } p_{\text{Total}}\left(x\right) > \tau, \\
+\infty, & \text{otherwise}.
\end{cases}
\label{eq:connect}
\end{equation}
Here, $I$ is the volume whose RCA mask is being connected, $I\left(x\right)$ is the voxel intensity at position $x$, $\mu$ and $\sigma$ are the mean and standard deviation of voxel intensities sampled from the reference volume along the centerline of the initial reference RCA mask, $p_{\text{Total}}$ is the RCA probability map of $I$ produced by TotalSegmentator, and $\tau$ is a threshold constant. By incorporating both intensity and anatomical priors, $C(x)$ encourages the minimum-cost path to stay within the anatomically plausible RCA region and discourages it from passing through adjacent cardiac structures. A tubular region centered along this path is merged with the proximal and distal fragments, yielding an RCA mask with a connected main trunk. 

\subsubsection{Regularize RCA Mask Shape}
\label{sec:regularize shape}
To regularize RCA mask shape, each RCA mask is then transformed into a sphere-union representation. Specifically, the RCA centerline is extracted using a 3D skeletonization algorithm~\cite{lee1994building}. For each centerline point $p$, a local vessel radius $r(p)$ is estimated as
\begin{equation}
r(p) = \mathbb{E}_{b \in S(p)} \| p - b \|_2 ,
\end{equation}
where
\begin{equation}
S(p) = \left\{ b \in B(M) \mid \ 
\arg\min_{p_0 \in P_{C}} \| p_0 - b \|_2 = p \right\}.
\end{equation}
Here, $M$ is the RCA mask, $B(M)$ is the boundary of $M$, and $P_{C}$ is the centerline. The centerline points together with their associated radii define a set of spherical regions. The union of these regions yields the RCA mask with regularized shape.

\subsubsection{Propagate RCA Mask Geometry}
The purpose here is to propagate the RCA mask geometry of the reference phase to non-reference phases. Due to motion artifacts, shape-regularized RCA masks of non-reference phases may still exhibit over-deformed centerlines, unreliable local radii, and missing or spurious branches. The shape-regularized RCA mask of the reference phase is generally more reliable. Propagating its geometry to non-reference phases can largely alleviate the remaining issues of the non-reference RCA masks.

Geometry propagation relies on deformation fields that roughly capture the RCA motion. These deformation fields are produced under the guidance of shape-regularized RCA masks using the proposed multi-resolution registration strategy. The same strategy is also employed in Stage 2 for capturing RCA motion using the final RCA masks produced by Stage 1, and is described in detail in Subsection~\ref{sec:stage2}.

Given the deformation fields, geometry propagation is performed for each non-reference phase. We first smooth the displacement along the reference RCA centerline to avoid overfitting to the potentially over-deformed centerline in the non-reference RCA mask. This is achieved by solving:
\begin{equation}
U^{\star}
=
\arg\min_{U}
\sum_{p}\left(
 \left\| U_p - D_p \right\|_2^2
+
\lambda \sum_{q\in \mathcal{N}\left(p\right)} \left\| U_p - U_q \right\|_2^2
\right),
\end{equation}
where 
$p$ is a point sampled from the reference RCA centerline, 
$D$ is the displacement from the reference phase to the non-reference phase and is derived from the deformation field,  
$\mathcal{N}\left(p\right)$ is the set of adjacent centerline points of $p$, and $\lambda$ is a weighting parameter. The optimized displacement $U^{\star}$ is then applied to the reference centerline to obtain the RCA centerline in the non-reference phase. The local RCA radii estimated in the reference phase are directly assigned to the propagated centerline in the non-reference phase, assuming negligible local radius variations across cardiac phases. The union of the resulting balls defined by the propagated centerline and radii yields the final RCA mask for the non-reference phase. 

The resulting non-reference RCA masks, together with the reference RCA mask obtained in Subsubsection~\ref{sec:regularize shape}, are used in Stage~2 for capturing RCA motion.

\subsection{Stage 2: Register the RCA}
\label{sec:stage2}

Stage 2 of CMGR focuses on capturing the RCA motion under the guidance of the plausible RCA masks produced by Stage 1. To this end, we propose a multi-resolution registration strategy built on Diffeomorphic Anatomical Registration using Exponentiated Lie algebra (DARTEL)~\cite{ASHBURNER200795}. The workflow of Stage 2 is illustrated in Fig.~\ref{fig:method}(b).
\subsubsection{DARTEL}
DARTEL registers a moving volume $m$ to a fixed volume $f$. The deformation field is modeled by a stationary velocity field (SVF) $u\left(x\right)$ through the following ordinary differential equation (ODE):
\begin{equation}
\frac{\partial\phi_s\left(x\right)}{\partial s}=u\left(\phi_s\left(x\right)\right),
\end{equation}
where $\phi_0$ is the identity transform, $\phi_S$ is the deformation field used to warp the moving volume $m$ to the fixed volume $f$, and $s\in[0,S]$ is the evolution time of the deformation field. This formulation ensures that $\phi_S$ is invertible. The SVF $u(x)$ is parameterized as a linear combination of first-degree B-spline basis functions and optimized by minimizing the following loss function:
\begin{equation}
\mathcal{L}\left(u\right)=\mathcal{L}_\text{D}\left(f,m\circ\phi_{S}\right)+\mathcal{L}_\text{D}\left(f\circ\phi_{S}^{-1},m\right)+\alpha \mathcal{L}_\text{R}\left(u\right),
\end{equation}
where $\phi_S$ and $\phi_S^{-1}$ are obtained by integrating $u$ and $-u$, respectively, using the scaling and squaring operation~\cite{dalca2019unsupervised}, $\mathcal{L}_\text{D}$ denotes the similarity loss, $\mathcal{L}_\text{R}$ denotes the regularization loss, and $\alpha$ is a weighting parameter. 
\subsubsection{Multi-Resolution Registration Strategy}
Directly applying DARTEL to register the reference volume to each non-reference volume may not capture RCA motion in the non-reference phases where the RCA exhibits large displacement or severe motion artifacts. To capture RCA motion for all non-reference phases, we propose a multi-resolution registration strategy guided by the plausible RCA masks.

The RCA mask guidance is introduced by preprocessing the volumes from the clinical multiphase reconstruction. For each volume, we highlight the region defined by its corresponding plausible RCA mask. Specifically, the RCA region is assigned a large constant intensity, making the RCA substantially brighter than other cardiac structures. The highlighted RCA encourages the subsequent registration to focus on aligning the RCA, and the preserved cardiac structures help keep the estimated deformation field compatible with whole-heart motion. In addition, because the plausible RCA masks generally exhibit reasonable geometry, this preprocessing also helps prevent the subsequent registration from distorting RCA geometry due to motion artifacts.

RCA motion is then captured by registering the RCA-highlighted reference volume to each RCA-highlighted non-reference volume in a multi-resolution manner. Both volumes are downsampled to multiple resolutions. Downsampling is performed using max-pooling to keep the highlighted RCA clearly identifiable at all resolution levels. The registration is performed by DARTEL for each resolution, and the SVF optimized for aligning lower resolution volumes is used to initialize the SVF for the next higher resolution level. In this multi-resolution strategy, lower resolution registration serves to capture global RCA motion, the SVF initialization enables the captured global RCA motion to be inherited at the next higher resolution level, and higher resolution registration serves to refine local RCA motion. Overall, this produces a deformation field that captures RCA motion between the full-resolution volumes.

Warping the original reference volume using the resulting deformation field produces an RCA-aligned volume for each non-reference phase. In the RCA-aligned volume, the RCA generally lies in a plausible position and has reasonable geometry. This volume is used in Stage 3 for capturing remaining cardiac motion.
\subsection{Stage 3: Produce Whole-Heart Deformation Fields}
The deformation field produced in Stage 2 generally captures RCA motion and preserves RCA geometry, but it may not sufficiently capture the motion of other cardiac structures. Stage 3 of CMGR aims to produce whole-heart deformation fields that preserve RCA motion and geometry, and better capture the motion of other cardiac structures. The workflow of Stage 3 is illustrated in Fig.~\ref{fig:method}(c).

For each non-reference phase, a deformation field is derived to capture the remaining cardiac motion without distorting the RCA due to motion artifacts. To be specific, the deformation field is designed to align non-RCA cardiac structures in the RCA-aligned volume from Stage 2 to those in the non-reference volume from the clinical multiphase reconstruction. The deformation field is also designed to preserve the position and geometry of the RCA in the RCA-aligned volume. To this end, both the RCA-aligned and non-reference volumes are preprocessed with our designed RCA-protection strategy. Specifically, an expanded RCA region is obtained by isotropically dilating the plausible non-reference RCA mask produced in Stage 1, and this region is assigned a large constant intensity in both volumes. The RCA protection strategy enables subsequent registration to establish strong correspondence between these protected regions, thereby discouraging RCA deformation. DARTEL is then applied to register the RCA-aligned volume after RCA protection to the RCA-protected non-reference volume, producing the deformation field that captures the remaining cardiac motion without RCA distortion.

The whole-heart deformation field $\phi_{t_n}$ is then obtained by:
\begin{equation}
\phi_{t_n} = \phi_{t_n}^{\left(\text{RCA}\right)}\circ \phi_{t_n}^{\left(\text{non-RCA}\right)}, 
\end{equation}
where $\phi_{t_n}^{\left(\text{RCA}\right)}$ is the RCA-motion-captured deformation field produced by Stage 2, $\phi_{t_n}^{\left(\text{non-RCA}\right)}$ is the deformation field that captures the remaining cardiac motion, and the composition order is compatible with (\ref{eq:warp}). The whole-heart deformation fields for all non-reference phases, together with the reference volume, are used in (\ref{eq:inter}) and (\ref{eq:warp}) to produce the motion-preserved, artifact-reduced, and continuous-time 4D cardiac CT sequence.

\begin{table*}[t]
\centering
\caption{Quantitative results in XCAT simulations. HD and MSD are in mm, and RMSE in HU. STD was computed across cardiac phases for each case and then averaged across cases. \textbf{Bold}: best; \underline{underlined}: second best.}
\label{tab:xcat}
\setlength{\tabcolsep}{1.02mm}{
\begin{tabular}{lcccccccccccc}
\toprule
&\multicolumn{6}{c}{Original Cardiac Phases}&\multicolumn{6}{c}{10$\times$ Denser Cardiac Phases}\\
\cmidrule(r){2-7}\cmidrule(r){8-13}
&\multicolumn{3}{c}{RCA}&\multicolumn{3}{c}{Entire Volume}&\multicolumn{3}{c}{RCA}&\multicolumn{3}{c}{Entire Volume}\\
\cmidrule(r){2-4}\cmidrule(r){5-7}\cmidrule(r){8-10}\cmidrule(r){11-13}
Method & Dice $\uparrow$ & HD  $\downarrow$ & MSD  $\downarrow$ & RMSE $\downarrow$ & SSIM $\uparrow$& LPIPS $\downarrow$ & Dice $\uparrow$ & HD  $\downarrow$ & MSD  $\downarrow$ & RMSE $\downarrow$ & SSIM $\uparrow$& LPIPS $\downarrow$\\
\midrule
FFD     & 0.48$\pm$0.33 & 9.30$\pm$7.64 & 2.99$\pm$2.84 & 36.3$\pm4.1$ & 0.91$\pm$0.01 & 0.20$\pm$0.03 &0.48$\pm$0.33&9.17$\pm$7.40&2.95$\pm$2.76&35.9$\pm$4.2& 0.91$\pm$0.01& 0.19$\pm$0.02 \\
DARTEL       & 0.55$\pm$0.31 & 8.65$\pm$8.46 & 2.48$\pm$2.78 & 35.4$\pm$3.3 & 0.91$\pm$0.01 & 0.20$\pm$0.03&0.56$\pm$0.31&8.59$\pm$8.19&2.43$\pm$2.68&34.9$\pm$3.3&0.91$\pm$0.01&0.19$\pm$0.02 \\
UGICON  & \underline{0.75$\pm$0.15} & \underline{3.39$\pm$3.41}& \underline{0.54$\pm$0.66} & \underline{34.9$\pm$2.3} & 0.91$\pm0.01$ & \textbf{0.18$\pm$0.02} &\underline{0.78$\pm$0.14}&\underline{3.30$\pm$3.20}&\underline{0.48$\pm$0.60}&\underline{34.1$\pm$2.1}&0.91$\pm$0.01&\textbf{0.18$\pm$0.02}  \\
Ours         & \textbf{0.78$\pm$0.11} & \textbf{1.91$\pm$0.86} & \textbf{0.34$\pm$0.16} & \textbf{34.4$\pm$2.1} & 0.91$\pm$0.01 & \textbf{0.18$\pm$0.01} &\textbf{0.82$\pm$0.08}&\textbf{1.77$\pm$0.76}&\textbf{0.28$\pm$0.12}&\textbf{33.0$\pm$1.6}&0.91$\pm$0.01&\textbf{0.18$\pm$0.01}  \\
\bottomrule
\end{tabular}}
\end{table*}

\begin{figure*}[t]
    \centering
    \includegraphics[width=1\linewidth]{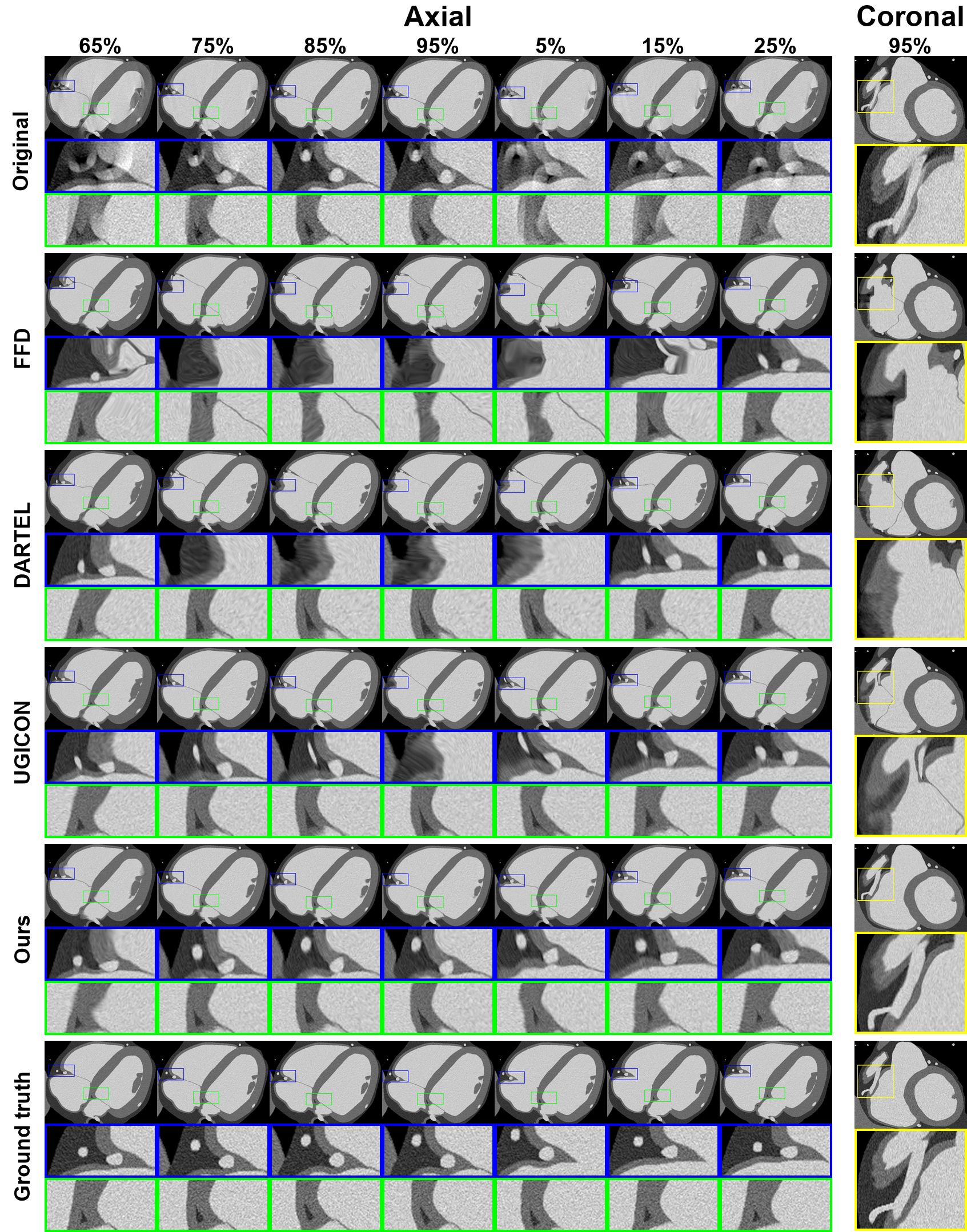}
    \caption{Visualization results for XCAT simulations. The cardiac phases close to the reference phase (45\%) were omitted from the visualization. The regions enclosed by blue, green and yellow boxes are magnified for visual clarity. The box positions may vary across cardiac phases and are fixed across the original multiphase reconstruction, all tested methods, and the ground truth. The display window is [-250, 450] HU. }
    \label{fig:xcat}
\end{figure*}
\begin{figure*}[t]
    \centering
    \includegraphics[width=1\linewidth]{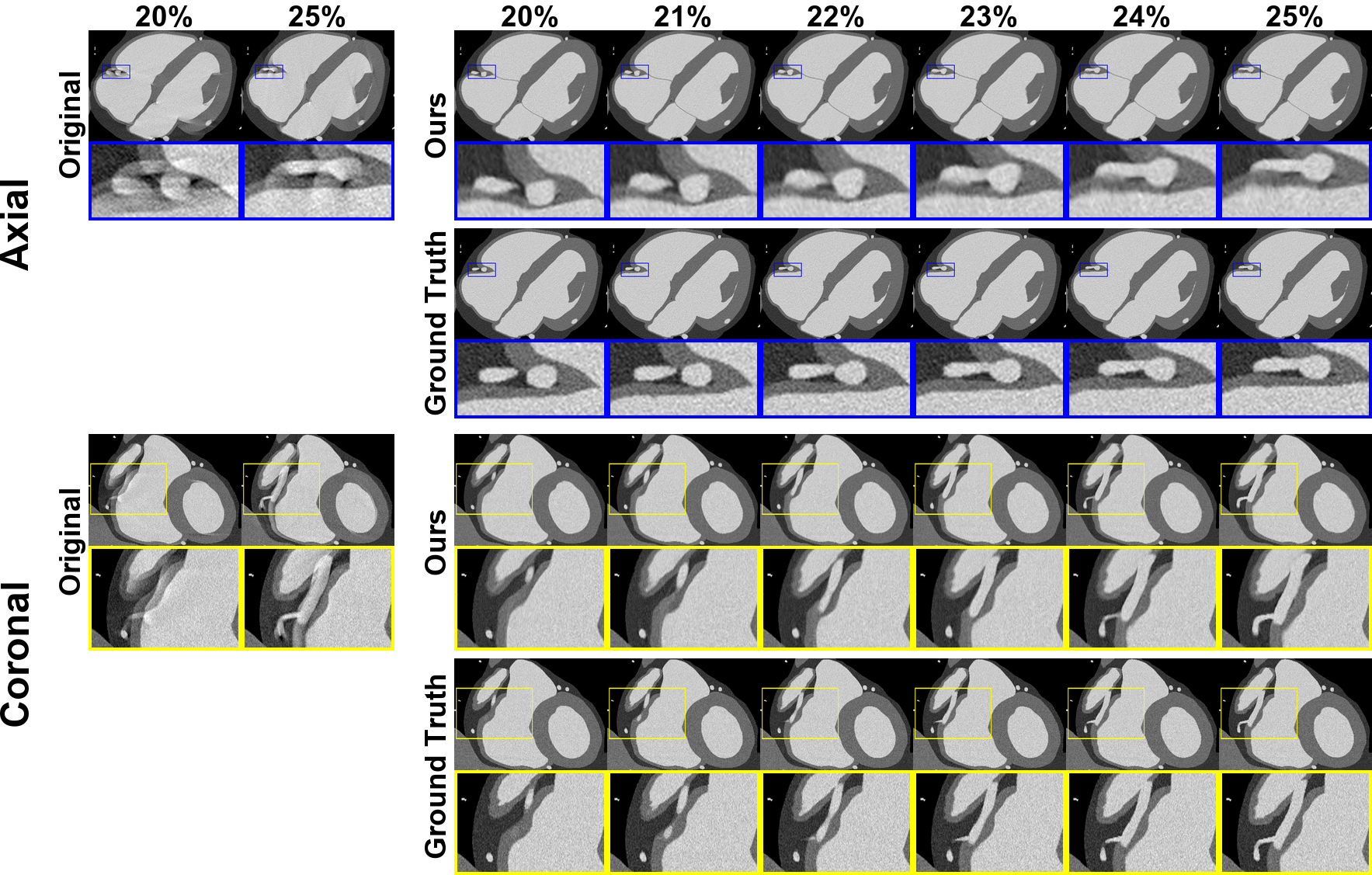}
    \caption{Visualization results for CMGR intermediate frames in XCAT simulations. The regions enclosed by blue and yellow boxes are magnified for visual clarity. The box positions are fixed for each view. The display window is [-250, 450] HU.}
    \label{fig:xcat_inter}
\end{figure*}

\begin{table}[t]
\centering
\caption{Quantitative results on clinical high-quality ED and ES pairs. HD and MSD are in mm, and RMSE in HU. STD was computed across pairs. \textbf{Bold}: best; \underline{underlined}: second best.}
\label{tab:clinical}
\setlength{\tabcolsep}{0.52mm}{
\begin{tabular}{lcccccc}
\toprule
&\multicolumn{3}{c}{RCA}&\multicolumn{3}{c}{Entire Volume}\\
\cmidrule(r){2-4}\cmidrule(r){5-7}
Method & Dice $\uparrow$ & HD  $\downarrow$ & MSD  $\downarrow$ & RMSE $\downarrow$ & SSIM $\uparrow$& LPIPS $\downarrow$ \\
\midrule
FFD     & 0.30$\pm$0.23 & 14.3$\pm$4.1 & 3.44$\pm$2.41 & 50.2$\pm4.4$ & 0.93$\pm$0.02 & 0.17$\pm$0.04  \\
DARTEL       & 0.45$\pm$0.28 & \underline{13.4$\pm$4.7} & 2.81$\pm$2.55 & 49.7$\pm$4.3 & 0.93$\pm$0.02 & 0.17$\pm$0.04\\
UGICON  & \underline{0.73$\pm$0.24} & 13.5$\pm$4.1& \underline{1.99$\pm$2.12} & \textbf{43.6$\pm$5.8} & \textbf{0.94$\pm$0.02} & 0.17$\pm$0.04  \\
Ours         & \textbf{0.83$\pm$0.04} & \textbf{8.6$\pm$3.4} & \textbf{0.33$\pm$0.09} & \underline{46.4$\pm$4.6} & 0.93$\pm$0.02 & 0.17$\pm$0.04  \\
\bottomrule
\end{tabular}}
\end{table}

\begin{table}[t]
\centering
\caption{FOR and NC in clinical data evaluation. STD was computed across cardiac phases for each case and then averaged across cases. \textbf{Bold}: best; \underline{underlined}: second best.}
\label{tab:clinical_for_nc}
\setlength{\tabcolsep}{2.62mm}{
\begin{tabular}{lcccc}
\toprule
&\multicolumn{2}{c}{Original Cardiac Phases}&\multicolumn{2}{c}{10$\times$ Denser Cardiac Phases}\\
\cmidrule(r){2-3}\cmidrule(r){4-5}
Method & FOR $\uparrow$ & NC  $\uparrow$ &FOR $\uparrow$ & NC $\uparrow$ \\
\midrule
FFD     & 0.74$\pm$0.05 & \underline{0.66$\pm$0.09} & 0.75$\pm$0.04 & \underline{0.66$\pm$0.08}   \\
DARTEL       & \underline{0.75$\pm$0.04} & 0.65$\pm$0.07 & \underline{0.76$\pm$0.03} & \underline{0.66$\pm$0.07} \\
UGICON  & 0.71$\pm$0.10 & 0.61$\pm$0.15& 0.73$\pm$0.08 & 0.63$\pm$0.13  \\
Ours         & \textbf{0.77$\pm$0.02} & \textbf{0.70$\pm$0.04} & \textbf{0.77$\pm$0.02} & \textbf{0.69$\pm$0.05}\\
\bottomrule
\end{tabular}}
\end{table}

\begin{figure*}[t]
    \centering
    \includegraphics[width=1\linewidth]{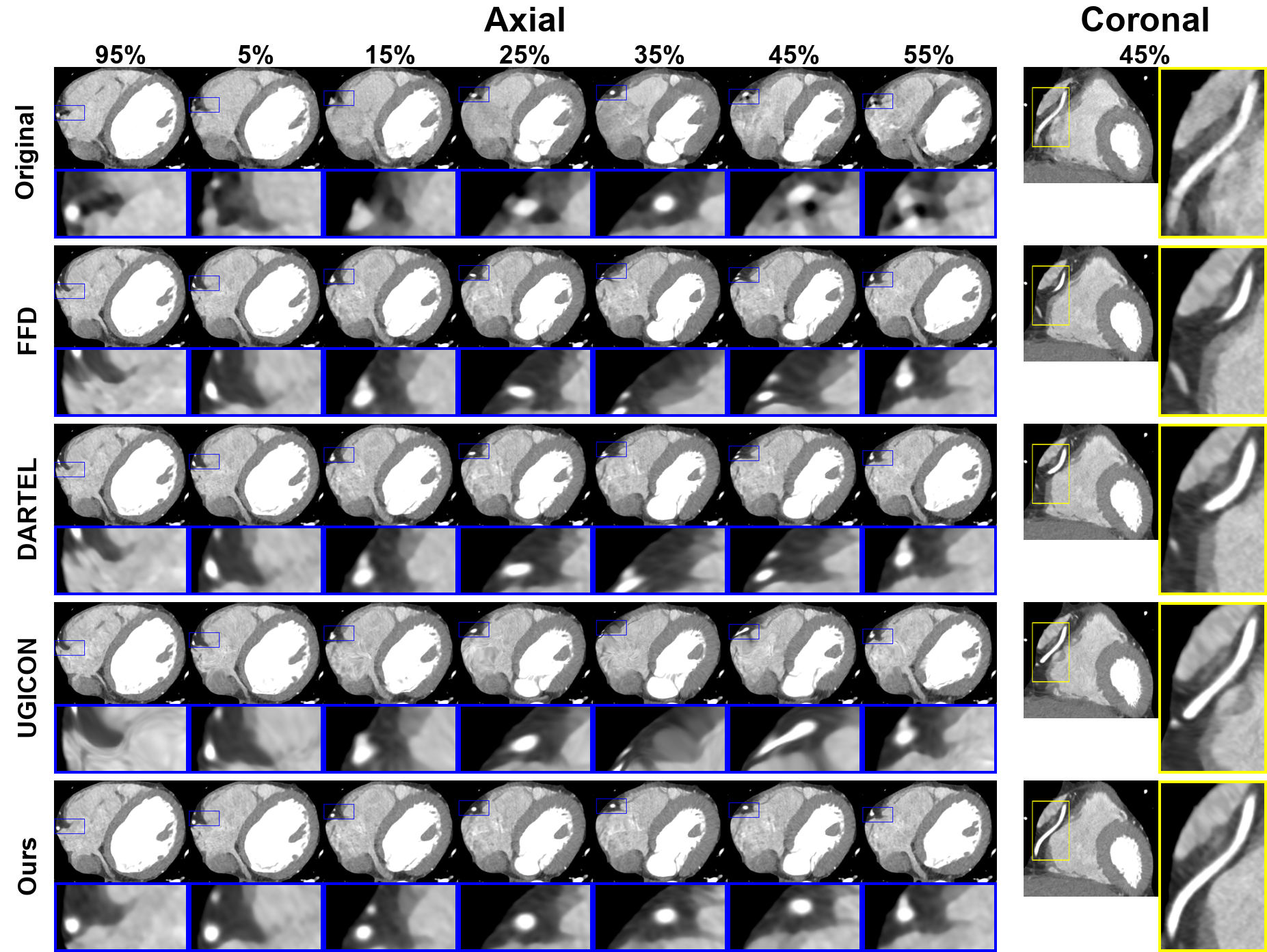}
    \caption{Visualization results for the first representative clinical case. The cardiac phases close to the reference phase (75\%) were omitted from the visualization. The regions enclosed by blue and yellow boxes are magnified for visual clarity. The box positions may vary across cardiac phases and are fixed across the original multiphase reconstruction and all tested methods. The display window is [-250, 450] HU.}
    \label{fig:clinical}
\end{figure*}
\begin{figure*}[t]
    \centering
    \includegraphics[width=1\linewidth]{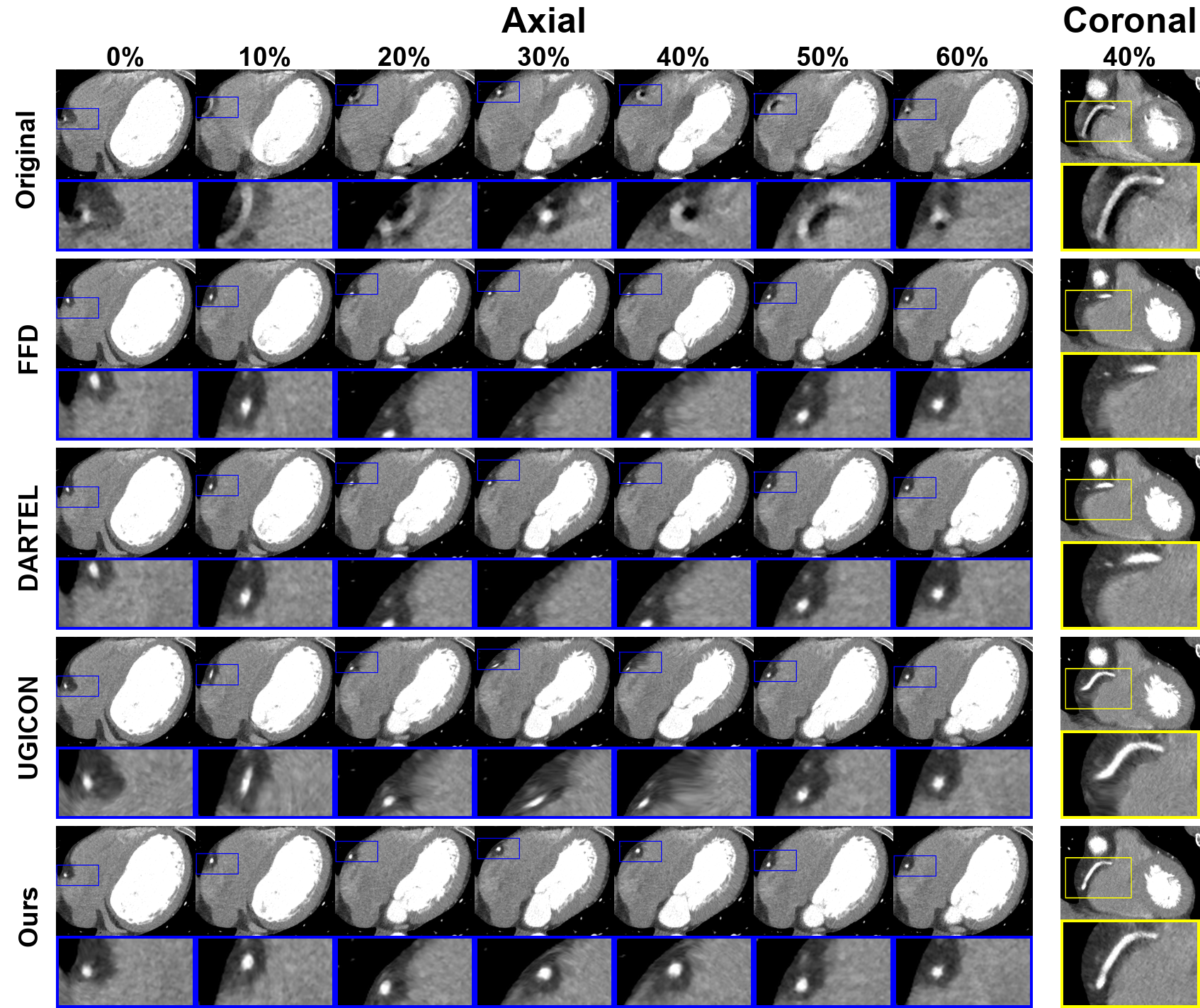}
    \caption{Visualization results for the second representative clinical case. The cardiac phases close to the reference phase (80\%) were omitted from the visualization. The regions enclosed by blue and yellow boxes are magnified for visual clarity. The box positions may vary across cardiac phases and are fixed across the original multiphase reconstruction and all tested methods. The display window is [-250, 450] HU.}
    \label{fig:clinical34}
\end{figure*}
\begin{figure*}[t]
    \centering
    \includegraphics[width=1\linewidth]{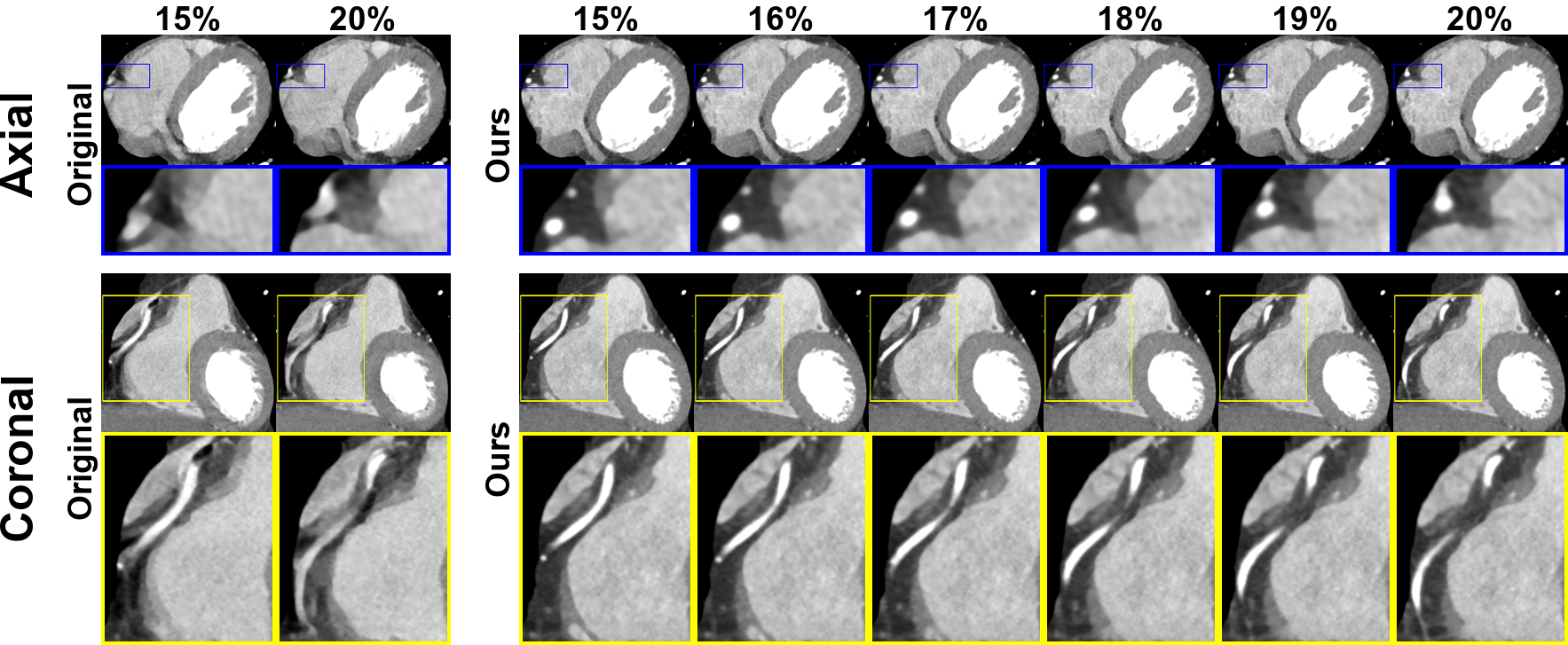}
    \caption{Visualization results for CMGR intermediate frames in clinical data evaluation. The regions enclosed by blue and yellow boxes are magnified for visual clarity. The box positions are fixed for each view. The display window is [-250, 450] HU.}
    \label{fig:clinical_inter}
\end{figure*}
\begin{figure}[t]
    \centering
    \includegraphics[width=1\linewidth]{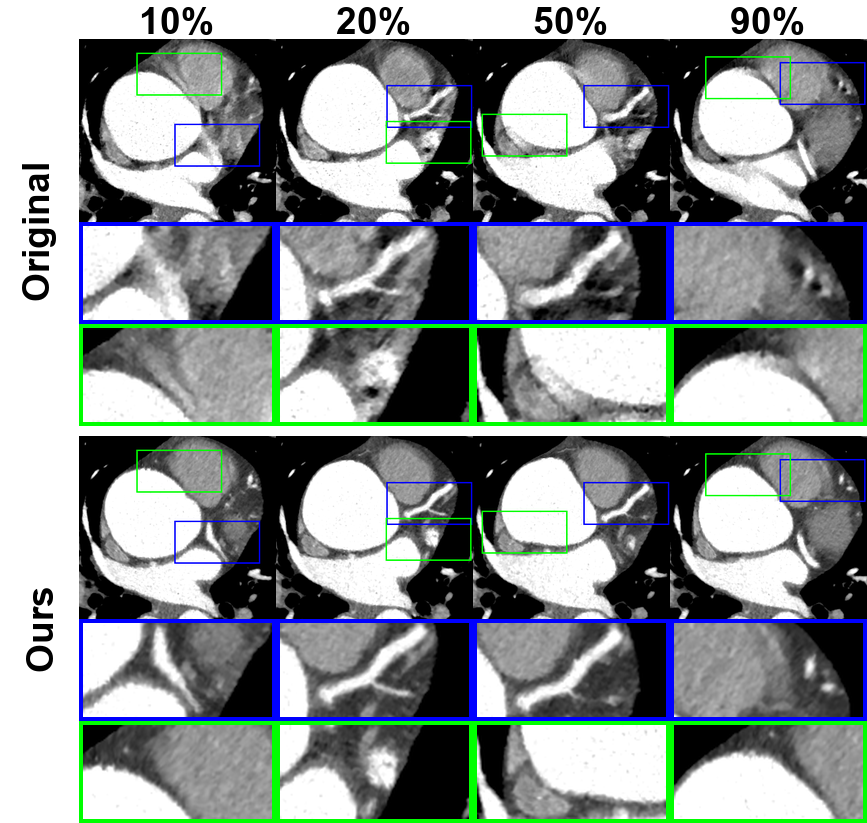}
    \caption{Visualization results for non-RCA cardiac structures in clinical data evaluation. The regions enclosed by blue and green boxes are magnified for visual clarity. The box positions may vary across cardiac phases and are fixed across the original multiphase reconstruction and CMGR. The display window is [-250, 450] HU.}
    \label{fig:clinical_nonrca}
\end{figure}

\begin{table*}[t]
\centering
\caption{Rank scores in the observer study. STD was computed across cases and observers. \textbf{Bold}: best; \underline{underlined}: second best.}
\label{tab:rank}
\setlength{\tabcolsep}{3.8mm}{
\begin{tabular}{lcccccccc}
\toprule
Rank Score $\uparrow$&\multicolumn{4}{c}{Radiologists}&\multicolumn{4}{c}{Researchers}\\
\cmidrule(r){2-5}\cmidrule(r){6-9}
&\multicolumn{2}{c}{RCA}&\multicolumn{2}{c}{Chamber \& Aorta}&\multicolumn{2}{c}{RCA}&\multicolumn{2}{c}{Chamber \& Aorta}\\
\cmidrule(r){2-3}\cmidrule(r){4-5}\cmidrule(r){6-7}\cmidrule(r){8-9}
Method & Motion & Geometry & Motion & Structure &  Motion & Geometry & Motion & Structure  \\
\midrule

DARTEL       & \underline{2.00$\pm$0.45} & \underline{1.98$\pm$0.42} & \underline{2.08$\pm$0.63} & \underline{2.08$\pm$0.56}& \underline{1.75$\pm$0.46} & \underline{1.99$\pm$0.38}&\underline{1.69$\pm$0.65}&\underline{2.17$\pm$0.64}\\
UGICON  & 1.24$\pm$0.59& 1.14$\pm$0.40& 1.38$\pm$0.60& 1.40$\pm$0.69& 1.35$\pm$0.60 & 1.09$\pm$0.29&1.64$\pm$0.79&1.21$\pm$0.55  \\
Ours         & \textbf{2.76$\pm$0.55}& \textbf{2.88$\pm$0.38} & \textbf{2.54$\pm$0.75}& \textbf{2.52$\pm$0.75} & \textbf{2.91$\pm$0.33}& \textbf{2.92$\pm$0.32}&\textbf{2.67$\pm$0.52}&\textbf{2.61$\pm$0.51} \\
\bottomrule
\end{tabular}}
\end{table*}

\begin{table}[t]
\centering
\caption{P-values in statistical tests of the observer study. Observer--case rank scores were used. Statistical tests were performed separately for radiologists and researchers, and larger p-values were reported.}
\label{tab:p-value}
\setlength{\tabcolsep}{1.57mm}{
\begin{tabular}{lcccc}
\toprule
P-value &\multicolumn{2}{c}{RCA}&\multicolumn{2}{c}{Chamber \& Aorta}\\
\cmidrule(r){2-3}\cmidrule(r){4-5}
Statistical Test& Motion & Geometry  &Motion & Structure \\
\midrule

Friedman      & $<0.001$ & $<0.001$& $<0.001$ & $<0.001$ \\
Wilcoxon(Ours$>$DARTEL) & $<0.001$ & $<0.001$& $0.019$ & $0.011$ \\
Wilcoxon(Ours$>$UGICON)        & $<0.001$ & $<0.001$& $<0.001$ & $<0.001$ \\
\bottomrule
\end{tabular}}
\end{table}
\section{Experiments}
\subsection{Data Preparation}
CMGR was evaluated using both XCAT and clinical data. In the XCAT simulations, we used both the male and female beating-heart representations of the 4D XCAT phantom~\cite{segars20104d}. The cardiac cycle was set to 1 s. Cone-beam CT acquisition was simulated over one complete cardiac cycle with a gantry rotation speed of 5 rotations per second. For each rotation, 800 views were acquired. The air photon count was set to $5\times10^5$. Reconstructions were performed at 20 cardiac phases with a temporal interval of 5\% of the cardiac cycle. For each phase, a $512\times512\times256$ volume with a voxel size of $0.3\times0.3\times0.5$ mm$^3$ was reconstructed using short-scan Feldkamp–Davis–Kress (FDK) followed by least-squares iterative reconstruction with 20 iterations. The resulting multiphase reconstructions were used as input to CMGR. The cardiac phase of 45\% was used as the reference phase.

For clinical data evaluation, we collected 16 cases from Fuwai Hospital and 9 cases from the Chinese People’s Liberation Army (PLA) General Hospital. The use of these data was approved by the institutional review boards (IRBs) of both hospitals (Approval No.: 2022-1787 and S2024-580). The data were acquired using GE and Siemens CT scanners. Each Fuwai case contained 16 to 20 phases at 5\% cardiac-cycle intervals, and each PLA case contained 10 phases at 10\% intervals. All clinical reconstructions were standardized to $512\times512\times256$. Volumes with more than 256 slices were center-cropped, and those with fewer slices were resampled. The in-plane pixel size ranged from 0.28 to 0.51 mm, and the slice thickness ranged from 0.40 to 0.63 mm. Among the 25 clinical cases, both ED and ES phases were of high image quality in 4 cases, and only the ED phase was of high quality in 17 cases. For these 21 cases, the ED phase was used as the reference phase. For the remaining 4 cases, only the ES phase was of high quality and was used as the reference phase.
\subsection{Implementation Details}
Most implementation settings of CMGR were fixed across all experiments. In RCA main trunk connection, the radius of the tubular region centered on the minimum-cost path was set to 1.5 mm. In geometry propagation, the weighting parameter $\lambda$ was set to 100. In DARTEL, the similarity loss $\mathcal{L}_\text{D}$ was the Mean Absolute Error, and the regularization loss $\mathcal{L}_\text{R}$ was the L2 norm of the spatial gradient of the SVF. In the multi-resolution registration strategy, six resolution levels with downsampling factors of 32, 16, 8, 4, 2, and 1 were used. In the RCA protection strategy, the RCA mask was isotropically dilated by 0.9 mm, and the constant intensity assigned to the protected RCA region was 2000 HU.

The threshold constant $\tau$ in (\ref{eq:connect}) for RCA main trunk connection, the regularization weight $\alpha$ in DARTEL, and the constant intensity assigned to the highlighted RCA region in the multi-resolution registration strategy were set differently for XCAT and clinical data. For XCAT, $\tau$ was set to $5\times10^{-7}$, $\alpha$ was set to 0.01, and the constant intensity was set to 2000 HU. For clinical data,  $\tau$ was set within the range $[0,2\times10^{-6}]$, $\alpha$ was set to 1 in Stages 1 and 2 and to 0.1 in Stage 3, and the constant intensity was set to 29000 HU.

\subsection{Evaluation Settings}
CMGR was compared with representative image-domain registration methods, including Nonrigid Registration using Free-Form Deformation (FFD)~\cite{rueckert1999nonrigid}, DARTEL~\cite{ASHBURNER200795}, and UniGradICON (UGICON)~\cite{tian2024unigradicon}. For each case, all tested methods used the same reference phase and registered the reference phase to each non-reference phase of the multiphase reconstruction. Using the resulting deformation fields and the reference volume, the continuous-time 4D cardiac CT sequence for each tested method was produced in the same way as CMGR, following (\ref{eq:inter}) and (\ref{eq:warp}). The comparison focuses on cardiac motion and structure.

\subsubsection{XCAT Evaluation} In XCAT evaluation, ground truth was available at all time points. Quantitative evaluation was performed in two temporal resolution settings. At lower temporal resolution, only volumes at the original cardiac phases of the multiphase reconstruction were evaluated. At higher temporal resolution, evaluation was performed at a set of cardiac phases that was 10 times denser than the original cardiac phases. Specifically, 9 intermediate phases between each pair of adjacent original phases were included. In both settings, Dice, Hausdorff Distance (HD), and Mean Surface Distance (MSD) were computed on the RCA masks to assess RCA motion capture, and Root Mean Square Error (RMSE), Structural Similarity Index Measure (SSIM), and Learned Perceptual Image Patch Similarity (LPIPS)~\cite{zhang2018perceptual} were calculated on the entire volumes to assess whole-heart motion capture. 
\subsubsection{Clinical Data Evaluation}
\paragraph{Quantitative Evaluation}
In clinical data evaluation, ground truth was generally unavailable. Therefore, ground-truth-dependent quantitative evaluation was restricted to four pairs of high-quality ED and ES phases. The ES phases were used as ground truth, and the ED phases were registered to the ES phases by the tested methods. Dice, HD and MSD were calculated on the RCA masks, and RMSE, SSIM and LPIPS were calculated on the entire volumes. This evaluates cardiac motion capture under low-motion-artifact conditions. 

In addition, Normalized Circularity (NC) and Fold Overlap Ratio (FOR)~\cite{ma2018evaluation} were computed on RCA masks for all 25 clinical cases to evaluate the circularity and symmetry of through-plane RCA cross-sections. The evaluation was performed at both the original cardiac phases and the 10$\times$ denser cardiac phases. 
\paragraph{Observer Study}
To further evaluate CMGR's performance in capturing cardiac motion and providing reasonable cardiac structures on clinical data, an observer study was performed on all 25 clinical cases. The study was conducted at the original cardiac phases without including intermediate frames. Two radiologists and three cardiac CT researchers were invited as the observers. The observers were provided with the clinical multiphase reconstruction and the corresponding sequences produced by DARTEL, UGICON, and CMGR. The three evaluated methods were randomly shuffled for each case and anonymized to the observers. The evaluation metrics were: (1) RCA Motion, assessing whether RCA motion aligns with that in the clinical multiphase reconstruction; (2) RCA Geometry, assessing whether RCA geometry is anatomically reasonable; (3) Chamber \& Aorta Motion, assessing whether the chamber and aorta motions align with those in the clinical multiphase reconstruction; and (4) Chamber \& Aorta Structure, assessing whether chamber and aorta structures are anatomically reasonable. For each case, each observer ranked the three anonymous methods as "best," "medium," and "worst" for each metric, and ties were not permitted.

To analyze the observer study results, we converted the rankings to rank scores. Specifically, "best" was converted to 3, "medium" to 2, and "worst" to 1. In addition, Friedman tests were performed to assess the differences among the evaluated methods, and one-sided Wilcoxon tests were performed to evaluate whether CMGR achieved statistically significant improvements over the other evaluated methods.

\section{Results}
\subsection{Results for XCAT Simulations}
\subsubsection{Quantitative Results}
In the XCAT simulations (Table~\ref{tab:xcat}), CMGR achieved the best performance in RCA-focused metrics. Compared to the other methods, CMGR achieved a 0.03 to 0.34 increase in Dice, a 44\% to 81\% decrease in HD, and a 37\% to 91\% decrease in MSD, and exhibited substantially smaller standard deviations (STDs) in HD and MSD. This demonstrates CMGR's superior capability and stability in capturing RCA motion.  In terms of the entire-volume metrics, CMGR achieved a 0.5 to 2.9 HU decrease in RMSE and similar SSIM and LPIPS, demonstrating CMGR's competitive performance in capturing whole-heart motion. In addition, the metrics of CMGR averaged across 10$\times$ denser cardiac phases were slightly better than those across original cardiac phases. This is because CMGR slightly overfitted to the motion-corrupted multiphase reconstructions at the original cardiac phases, and intermediate frames averaged out such overfitting to some extent. These results demonstrate that intermediate CMGR frames generally provide plausible transitions between CMGR volumes at adjacent original cardiac phases.

\subsubsection{Visualization Results}
Visualization results for XCAT simulations are shown in Fig.~\ref{fig:xcat}. Compared to the original multiphase reconstruction, CMGR substantially alleviates motion artifacts in both the RCA and the left ventricle (LV) corner. Compared to the other methods, CMGR better aligns with the ground truth in both RCA position and RCA shape in axial and coronal views. This demonstrates CMGR's superior capability in capturing RCA motion and providing reasonable RCA shape. In addition, the LV corner in CMGR generally aligns with that in the ground truth, indicating CMGR's competitive performance in capturing whole-heart motion.

Visualization results for CMGR intermediate frames in XCAT simulations are shown in Fig.~\ref{fig:xcat_inter}. At cardiac phases 20\% and 25\%, CMGR substantially alleviates motion artifacts in the RCA compared to the original multiphase reconstruction. Furthermore, CMGR provides the intermediate frames that are unavailable in the original multiphase reconstruction. In axial and coronal views, the intermediate CMGR frames provide a plausible transition between cardiac phases 20\% and 25\% and generally align with the corresponding ground-truth frames.
\subsection{Results for Clinical Data Evaluations}
\subsubsection{Quantitative Results}
In the evaluation on clinical high-quality ED and ES pairs (Table~\ref{tab:clinical}), CMGR achieved the best performance in RCA-focused metrics. Compared to the other methods, CMGR achieved a 0.10 to 0.53 increase in Dice, a 36\% to 40\% decrease in HD, and an 83\% to 90\% decrease in MSD, and exhibited substantially smaller STDs in Dice and MSD. This demonstrates CMGR's superior capability and stability in capturing RCA motion under low-motion-artifact conditions. In terms of the entire-volume metrics, CMGR achieved the second-best RMSE and similar SSIM and LPIPS, demonstrating CMGR's competitive performance in capturing whole-heart motion.

In the clinical data evaluation on RCA shape (Table~\ref{tab:clinical_for_nc}), CMGR achieved the best FOR and NC and exhibited the smallest STDs for both metrics. This demonstrates that CMGR provides more reasonable and stable RCA shapes compared to the other methods. In addition, the FOR and NC of CMGR averaged across 10$\times$ denser cardiac phases were comparable to those across original cardiac phases, showing that intermediate CMGR frames generally provide reasonable RCA shapes.

\subsubsection{Visualization Results}
Visualization results on clinical data are shown in Fig.~\ref{fig:clinical} and Fig.~\ref{fig:clinical34}. Compared to the original multiphase reconstruction, CMGR substantially alleviates motion artifacts in RCA. Compared to the other methods, CMGR demonstrates superior capability and stability in capturing RCA motion and providing reasonable RCA shape. This superiority is demonstrated in three aspects. First, CMGR provides more plausible RCA positions in the axial view at multiple cardiac phases, including phases 35\% and 95\% in Fig.~\ref{fig:clinical}, and 40\% in Fig.~\ref{fig:clinical34}. Second, in the axial view, the comparison methods exhibit noticeable RCA shape distortion at some phases, especially UGICON at phase 45\% in Fig.~\ref{fig:clinical} and  10\% and 30\% in Fig.~\ref{fig:clinical34}. In contrast, CMGR provides reasonable and stable RCA shape across cardiac phases. Third, CMGR provides more plausible RCA anatomical courses in the coronal view compared to the other methods.

Visualization results for CMGR intermediate frames on clinical data are shown in Fig.~\ref{fig:clinical_inter}. At cardiac phases 15\% and 20\%, CMGR substantially alleviates motion artifacts in the RCA. Furthermore, the intermediate CMGR frames provide a plausible transition between cardiac phases 15\% and 20\% in axial and coronal views.

Visualization results for non-RCA cardiac structures in clinical data evaluation are shown in Fig.~\ref{fig:clinical_nonrca}. Compared to the original multiphase reconstruction, CMGR alleviates motion artifacts in the left coronary artery, right ventricle, left atrium, and aorta. This demonstrates that CMGR can reduce motion artifacts in cardiac structures besides the RCA.

\subsubsection{Observer Study Results}
In the observer study (Tables~\ref{tab:rank} and~\ref{tab:p-value}), CMGR achieved the best performance in both radiologist and researcher evaluations across all four metrics. In RCA Motion and RCA Geometry, CMGR achieved the best rank scores with p-values less than 0.001. This demonstrates CMGR's superior capability in capturing RCA motion and providing reasonable RCA geometry compared with the other evaluated methods, and the superiority in RCA is statistically significant. In Chamber \& Aorta Motion and Chamber \& Aorta Structures, CMGR achieved the best rank scores with p-values less than 0.02. This statistically significant improvement in chamber and aorta metrics may be associated with the more plausible RCA position in CMGR, which may contribute to more reasonable deformation of adjacent cardiac chambers. This result demonstrates CMGR's competitive performance in capturing whole-heart motion and providing reasonable cardiac structures besides the RCA.

\subsection{Summary for Results}
The results show CMGR's capability in producing motion-preserved, artifact-reduced, and continuous-time cardiac CT sequences from multiphase reconstructions without access to raw projection data. For motion preservation, compared with the other methods, CMGR shows superior capability and stability in capturing RCA motion and providing reasonable RCA shape, and shows competitive performance in capturing whole-heart motion. For artifact reduction, CMGR reduces motion artifacts compared with the original multiphase reconstruction in both the RCA and other cardiac structures. For time continuity, CMGR intermediate frames generally maintain reasonable RCA shapes and provide plausible transitions between CMGR volumes at adjacent original cardiac phases. 

\section{Conclusion}
In conclusion, CMGR provides an effective approach for constructing a 4D cardiac CT dataset that is generally suitable to serve as pseudo ground truth for 4D cardiac CT imaging research. Compared with previous pseudo-ground-truth construction strategies~\cite{lossau2019motion,deng2023tt,segars20104d}, CMGR is able to provide patient-specific cardiac anatomy and cardiac motion with a generally acceptable motion artifact level. Since CMGR is able to produce continuous-time cardiac CT sequences without access to raw projection data, CMGR is generally more suitable for dataset construction than cardiac CT reconstruction methods~\cite{maier2021deep,deng2026marvel}. In the comparison with representative image-domain registration methods, CMGR demonstrates superior capability and stability in capturing RCA motion and providing reasonable RCA shape, and shows competitive performance in capturing whole-heart motion. We believe that CMGR has the potential to facilitate cardiac CT system design and reconstruction algorithm development by providing pseudo-ground-truth datasets.

\section{Limitations and Future Work}
CMGR has two limitations. First, CMGR is not capable of capturing the opening and closing motion of cardiac valves. Specifically, a valve in the CMGR-produced sequence may remain open or closed throughout the cardiac cycle, with its status largely determined by the reference phase. This limitation arises because Stage 3 of CMGR may be unable to capture the motion of valve leaflets, as the leaflets are anatomically thin and often exhibit motion artifacts or large displacements. In addition, valve opening and closing motion may not be invertible in the image domain, where the valve leaflets appear split when open and merged when closed. This also hinders CMGR from capturing the valve motion.

Second, for the non-reference phase where RCA exhibits significant large displacement relative to the reference phase, CMGR may slightly over-compress or over-stretch the cardiac structures adjacent to the RCA. This limitation arises from Stage 2 of CMGR, where SVF is used to model the deformation field in RCA registration. SVF assumes the velocity at a spatial point to be independent of the evolution time of the deformation field. Therefore, when adjacent cardiac structures pass through the RCA trajectory, they share the same velocity as the RCA. This can lead to slight abnormal deformation of adjacent cardiac structures if the RCA trajectory is substantially long. This limitation may not be alleviated in Stage 3, since further deformation in an expanded RCA region is discouraged by the RCA protection strategy.

In addition to addressing the above limitations, we plan to apply CMGR to more patient data to construct a large-scale 4D cardiac CT dataset. We also plan to use the dataset in developing generative-model-based 4D cardiac CT reconstruction algorithms and in simulations for stationary CT systems.

\section*{Acknowledgment}
This work was supported by the National Natural Science Foundation of China (No. 12327809 and No. 62031020).
\end{document}